\documentclass[twocolumn,showpacs,preprintnumbers,amsmath,amssymb,prb]{revtex4-1}

\usepackage{graphicx}
\usepackage{dcolumn}
\usepackage{bm}
\usepackage{psfrag}
\usepackage{color}
\usepackage{subfigure}
\usepackage{dsfont}

\begin{document}
\title{Electronic topological transitions in Cd at high pressures}

\author{Varadharajan Srinivasan$^{(1)}$, B.K. Godwal$^{(3)}$, Jeffrey C. Grossman$^{(2)}$, and Raymond Jeanloz$^{(3)}$}
\affiliation{(1) Department of Chemistry, Indian Institute of Science Education and Research Bhopal, Bhopal 462 066, Madhya Pradesh, India.}
\affiliation{(2) Department of Materials Science and Engineering, Massachusetts Institute of Technology, Cambridge, MA 02145.}
\affiliation{(3) Earth and Planetary Sciences, University of California, Berkeley, CA 94720.}

\date{\today}

\begin{abstract}
Pressure-induced changes in the Fermi surface of Cd up to 40 GPa are studied using highly accurate density-functional theory calculations. The topology of Fermi surface changes at pressures of 2, 8, 12 , 18 and 28 GPa indicating electronic topological transitions (ETTs). Structural parameters, compressibility data and elastic constant reveal anomalies across these ETTs. The computed equation of state at 300 K is in excellent agreement with the experimental data.  In view of the highly controversial nature of these ETTs the present studies are aimed at motivating the experimentalists for their direct detection at high pressures.
\end{abstract}

\maketitle

\section{Introduction}

Predictions of phase transitions, resolving controversies in the experimental data   and providing the underlying mechanism are some of the major goals of  condensed matter  theory~\cite{toledano_symmetry_2009}. The subtle changes in electronic band structure can cause major changes in chemical and physical properties of certain elements and compounds, for example small variations in composition causing changes in Fermi surface topology have a big impact on the material. Some of these anomalies in material properties are caused by the proximity of various energy band extrema   to the Fermi energy and occur due to their passage through it with variation in temperature or alloying, widely known as a Lifshitz or electronic topological transition (ETT)~\cite{Lifshitz_1960}. Alloying and temperature can lead to changes in band filling but may smear the electronic states and hence the observable effects. The use of pressure  is a more direct way to compare against first principles theory. Recent high pressure angle dispersive x-ray diffraction studies using a high resolution synchrotron source indicate that there is evidence in the compressibility data for the changes predicted by {\it ab initio} theories due to changes in band topology~\cite{speziale_axial_2008}. The major effects observed by varying the composition can also be validly interpreted as a result of subtle electronic changes widely known as ETT effects.

Elemental solids Zn and Cd, which crystallize in the hcp structure and have large values of axial ratio (c/a) compared to the ideal value (1.633), have been the subject of intense experimental and theoretical investigations~\cite{lynch_effect_1965, kenichi_zn_1995, kenichi_structural_1997,  meenakshi_distorted_1992, fast_anomaly_1997, godwal_anomalies_1997, novikov_anisotropy_1999, li_phonon_2000, steinle-neumann_absence_2001, qiu_structural_2004, sinko_effect_2005,  potzel_electronically_1995, morgan_inelastic_1996, klotz_is_1998, olijnyk_pressure_2000,schulte_effect_1996,kenichi_absence_1999,kenichi_axial_2002}. Their large c/a values (1.856 for Zn and 1.89 for Cd) lead to anisotropy in Fermi surface topology, transport and other physical properties. Bondenstedt {\it et al.}~\cite{bodenstat_1986} have argued that these anisotropies are due to asymmetric charge distribution of the {\it p} electrons. This makes the Fermi surface of Cd and Zn different from normal hcp metals with ideal c/a ratio and gives rise to a giant Kohn anomaly. Experiments using ethanol-methanol as a pressure medium~\cite{lynch_effect_1965, kenichi_zn_1995, kenichi_structural_1997, schulte_effect_1996} exhibited an anomalous departure from a smooth c/a decrease for Zn near 10 GPa and for Cd near 8 GPa. This anomaly in Zn was, however, contradicted with measurements using helium as a pressure medium where hydrostatic pressures are ensured throughout the measurement regime~\cite{kenichi_absence_1999}. On the theoretical side, earlier calculations which supported a $c/a$ anomaly in Zn and associated it with an electronic phase transition~\cite{meenakshi_distorted_1992, fast_anomaly_1997, godwal_anomalies_1997, novikov_anisotropy_1999,li_phonon_2000} were challenged based on the absence of such an anomaly in electronic band structure calculations using dense k-point meshes for the Brillouin zone (BZ) summations~\cite{steinle-neumann_absence_2001}. However, the doubts about the existence  of an anomaly in the pressure variation of axial ratio  in Zn has been cleared by Qiu {\it et al}~\cite{qiu_structural_2004} who find the anomaly in Zn with 530 as well as 5300 k-points in the irreducible wedge of the Brillouin zone. Further, these studies also observe anomalies in the ratio of linear compressibilities   and   elastic moduli over large pressure ranges for both Zn and Cd. Based on a combination of detailed theoretical studies and a variety of experimental measurements ranging from Mossbauer~\cite{potzel_electronically_1995}, x-ray diffraction~\cite{kenichi_zn_1995,kenichi_structural_1997,kenichi_axial_2002}, and neutron scattering~\cite{morgan_inelastic_1996,klotz_is_1998} to Raman studies~\cite{olijnyk_pressure_2000} on Zn,  there appears to be agreement that structural anomalies due to ETT in Zn are less pronounced although they exist. Their small magnitude is perhaps within the experimental uncertainty making it difficult to resolve them by x-ray diffraction at high pressures and ambient temperature. The important conclusion of the studies by Qiu {\it et al}~\cite{qiu_structural_2004} is the finding of a stronger structural anomaly in Cd compared to Zn. In view of such results from past studies we have focused our studies on Cd.

\begin{figure}[t!]
\begin{center}
\includegraphics[width=3.5in,clip=]{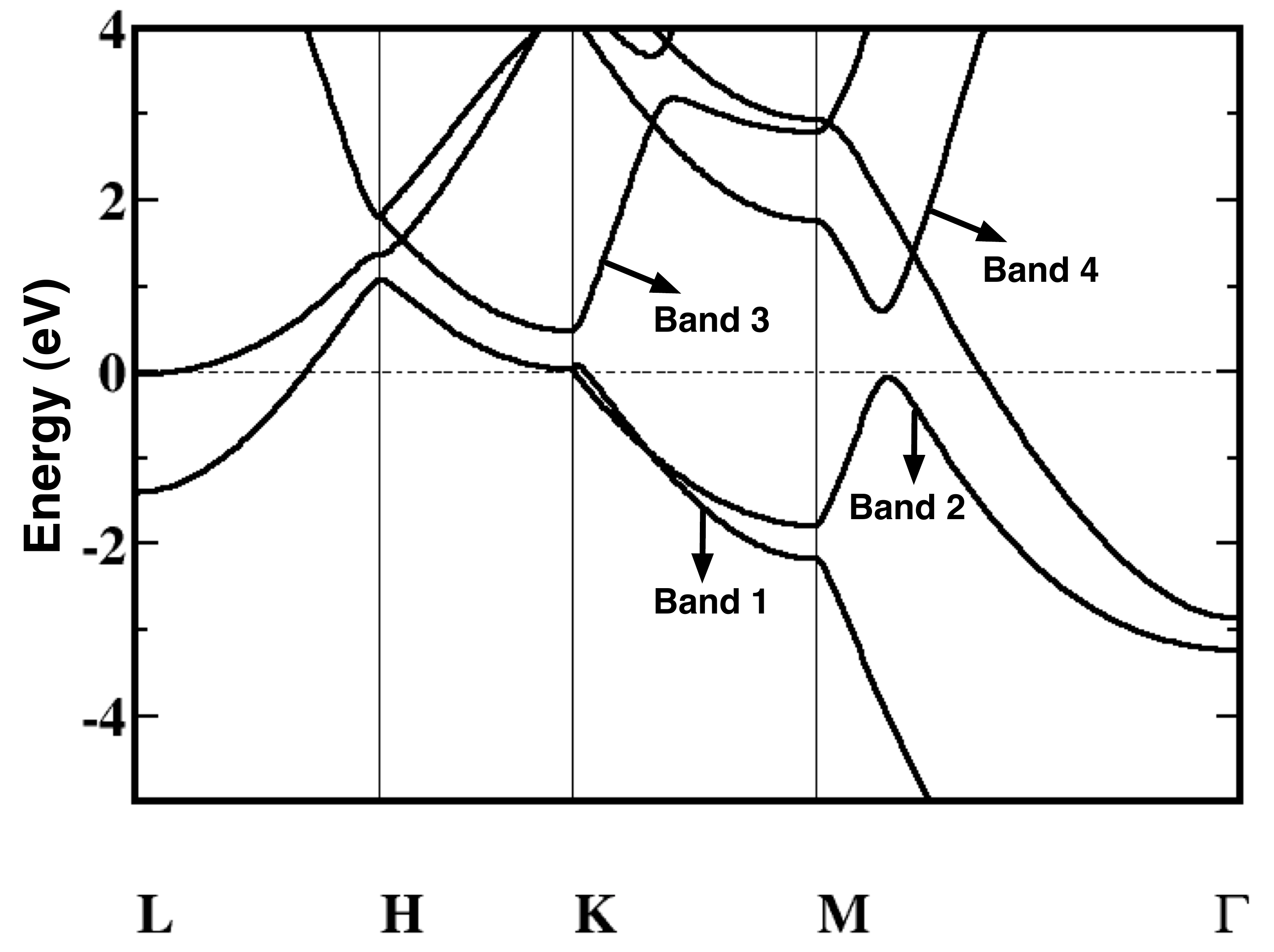}
\end{center}
\caption{Band structure of Cd showing the 4 bands whose crossing the Fermi surface gives rise to a total of 5 ETTs.}
\label{fig:bands}
\end{figure}

In this study we investigate the structural behavior of Cd over a wide pressure range of 0 to 40 GPa.  We have employed highly accurate density-functional theory (DFT) techniques to predict the variation of lattice parameters with pressure, obtaining an excellent agreement of our calculated equation of state with the experiment. Electronic phase transition (ETT) as a true phase transition is expected to appear only at T=0 K in highly pure samples and otherwise is just a crossover. All our calculations are thus at 0 K. In contrast to previous studies~\cite{qiu_structural_2004},  our calculations extend to higher pressures to the order of 40 GPa. Consequently, we also observed 3 new anomalies in addition to those already reported. As the subtle electronic structure changes under pressure are central to the association of ETT with observed structural anomalies, we have computed the Fermi surface (FS) at different pressures, identifying key topological changes. We show that these changes correlate well with the predicted anomalies in the lattice parameters as well as elastic constants. The co-location of the electronic and structural anomalies provide strong evidence in favor of ETTs as being the origin of the experimentally observed anomalies in Cd.

\section{Methodology}

Our DFT calculations used a generalized-gradient approximation for the exchange-correlation (XC) functional in the Perdew, Burke and Ernzerhof  scheme~\cite{perdew_generalized_1996} (PBE). This choice was motivated by previous studies~\cite{novikov_lda_1997} indicating that the local density approximation to the XC displays several failures in describing ETTs which are corrected satisfactorily by PBE. We employed the QUANTUM-ESPRESSO code~\cite{giannozzi_quantum_2009}  which uses plane-wave basis sets to represent wave-functions and charge-densities. The ionic core of Cd was modeled using an ultra-soft pseudopotential~\cite{vanderbilt_soft_1990} generated in the $4d^{10}5s^{2}5p^{0}$ configuration including a non-linear core-correction scheme to account for non-linearities in the XC functional. A highly dense uniform $24\times24\times24$ k-point mesh was used to sample the whole BZ. Full structural optimizations at constant pressure, yielding equilibrium volumes as well as structural parameters, were performed using the Wentzcovich variable-cell scheme~\cite{wentzcovitch_invariant_1991}. While charge-densities and energies were well-converged at a PW cut-off of 27 Ry, the variable-cell scheme relies on accurate stress computations which often require a   higher kinetic energy cut-off. Careful convergence tests indicated that structures optimized at the above cut-off yielded stresses within 1-2 kbar of the fully-converged values at 50 Ry and thus were used as such for further calculations below. Cell optimizations were performed in the hexagonal symmetry for pressures ranging from 0 to 40 GPa in order to obtain pressure-volume curves comparable with the measured experimental range.  On each of the resulting structures we have further computed band-structure and Fermi surfaces using similar calculation parameters. The elastic constant  C$_{33}$  was  also computed and its convergence with the parameters ensured. Our approach differs from that of Qiu {et al}~\cite{qiu_structural_2004} in the use of the variable-cell method to directly obtain optimized cell parameters at a given pressure instead of generating free-energy curves at various pressures along the epitaxial Bain path. 

\section{Results and Discussion}

\begin{figure*}[t!]
\begin{center}
\includegraphics[width=7.0in,clip=]{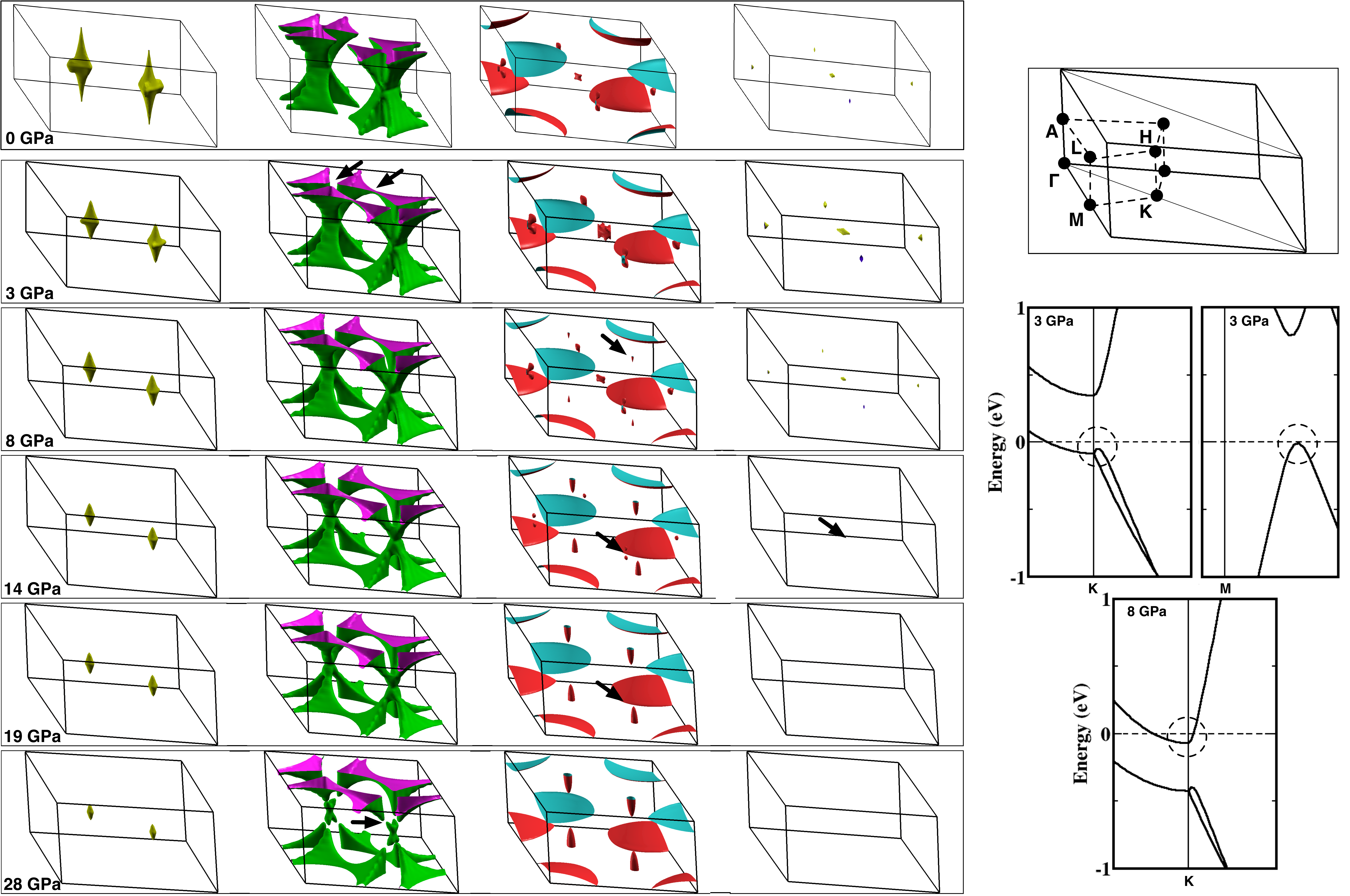}
\end{center}
\caption{Fermi surface contributions from the 4 bands of interest at pressures  just above the ETTs (left). The surfaces are plotted in the regular hexagonal reciprocal lattice unit cell (shown on top right) rotated to enable a clear view of the topological changes at each ETT. The origins of the first and second transitions in terms of bands crossing the Fermi level at 3 and 8 GPa, respectively are also shown (middle and bottom right panels).}
\label{fig:fermi}
\end{figure*}

\subsection{Band structure and Fermi surface topology}

\begin{figure*}[t!]
\begin{center}
\includegraphics[width=5.0in,clip=]{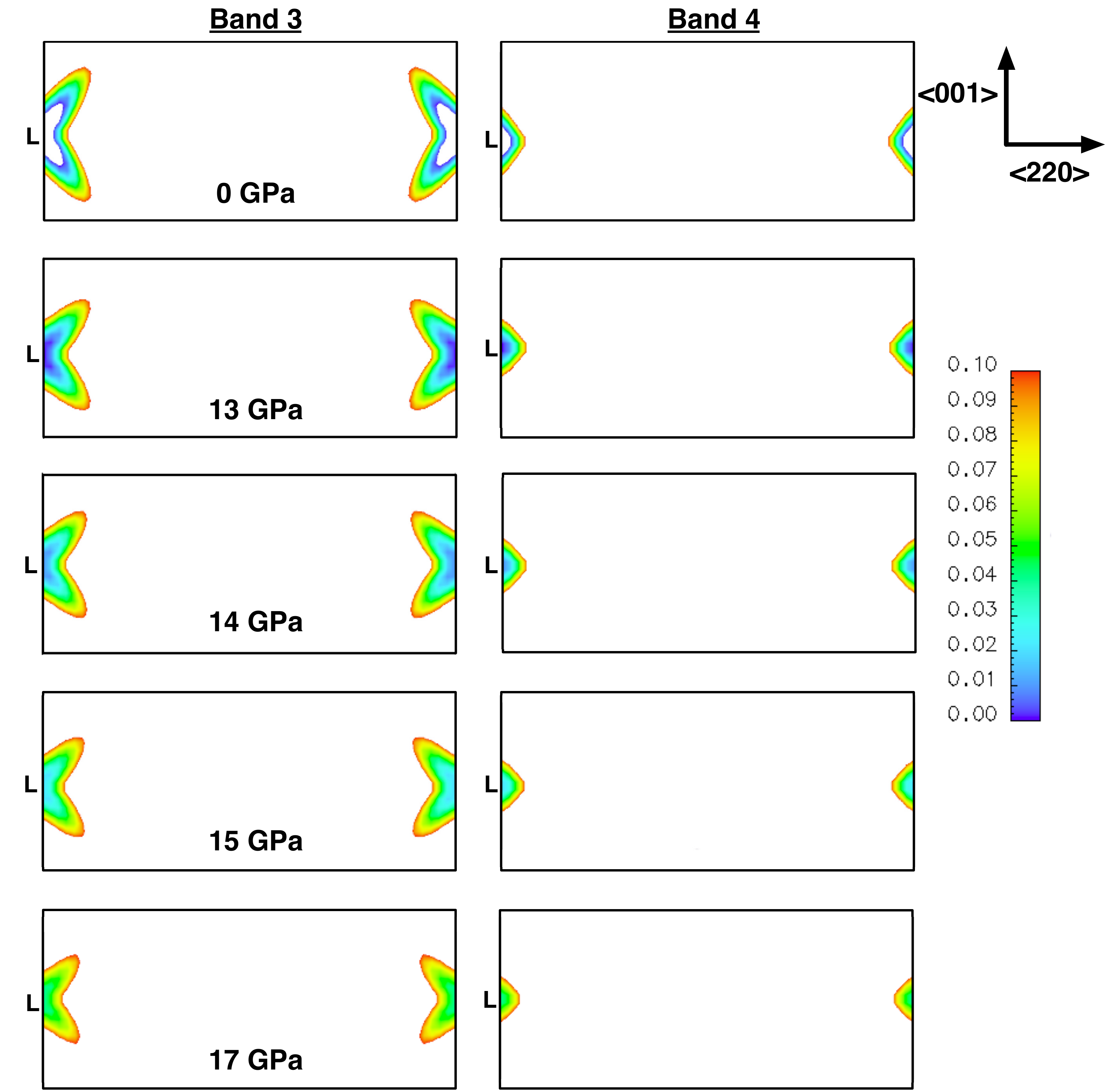}
\end{center}
\caption{Projection of the Fermi surface contribution from bands 3 and 4 on the $\left[220\right]$ plane at 0, 13, 14, 15 and 17 GPa. The butterfly-shaped contribution to the Fermi-surface around the L point from band 3 (left) as well as the diamond-shaped contribution from band 4 (right) undergo topological changes between 13 and 14 GPa corresponding to the third transition (ETT 3). The wings of the butterfly (see text) in band 3 persist beyond 14 GPa but finally vanish around 17 - 18 GPa giving rise to the fourth transition (ETT 4). }
\label{fig:fs_2d}
\end{figure*}

\begin{figure*}[t!]
\begin{center}
\includegraphics[width=3.5in,clip=]{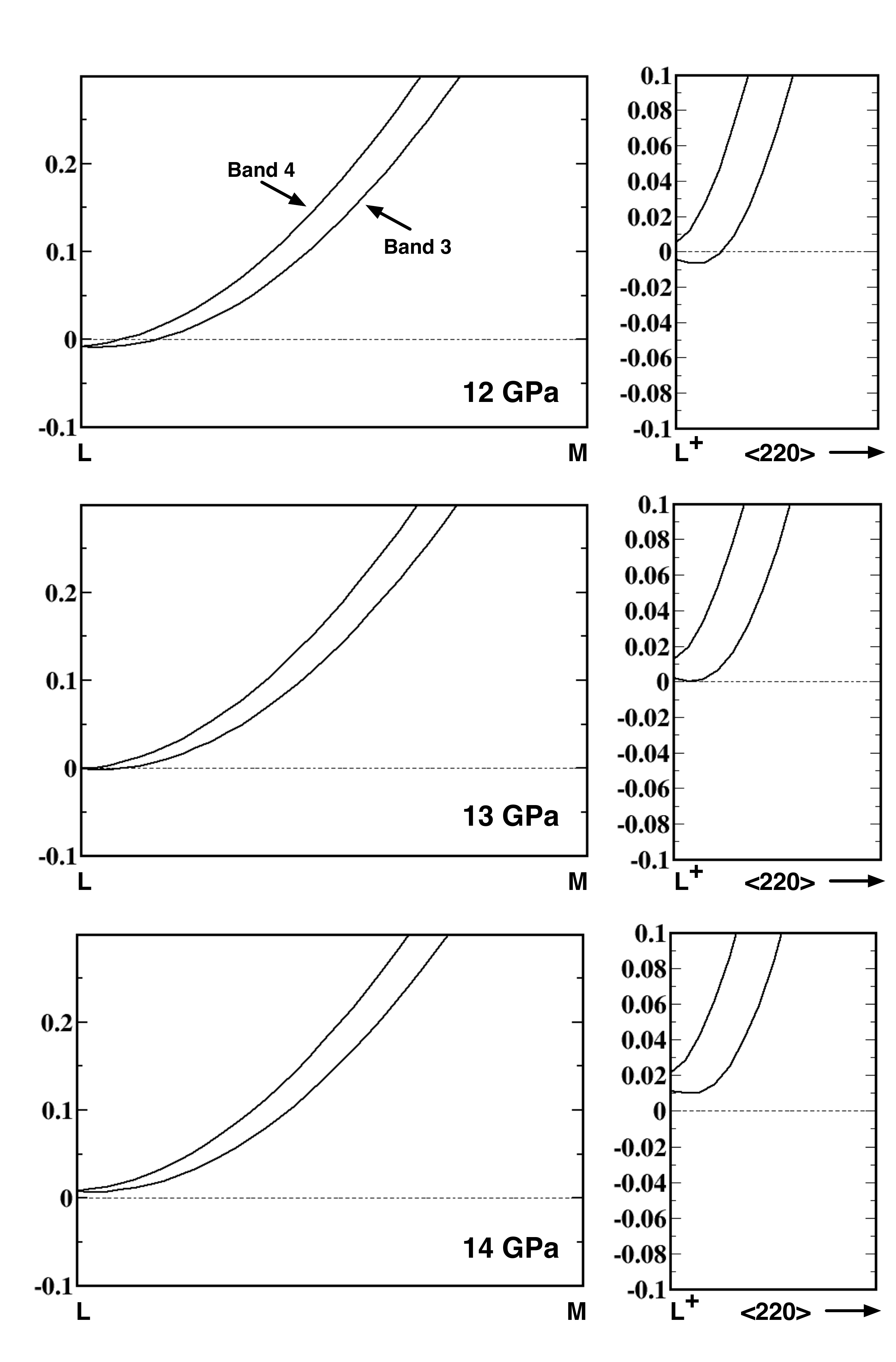}
\end{center}
\caption{ Left panels illustrate the occurrence of ETT 3 where both bands 3 and 4 (plotted along LM) move across the Fermi surface around point L. In the right panels the two bands are plotted along a line in the $\left[220\right]$ plane slight above the L point ($L^+$). It can be seen that band 3 which is roughly parabolic along LM develops a negative curvature at $L^{+}$ resulting in the band crossing the Femi surface twice, hence, giving rise to the ``wings of the butterfly" referred to in Fig.~\ref{fig:fs_2d}. As we move away from $\left[220\right]$ parallel to this line the wing feature persists even for pressures higher than 13 GPa finally vanishing around 17 - 18 GPa.}
\label{fig:ett34_1d}
\end{figure*}

Figure~\ref{fig:bands} shows the band-structure along a few high symmetry directions in bulk Cd at  ambient pressure. The Fermi surface is essentially composed of contributions from 4 bands. It is the variation in dispersion around the Fermi energy of these bands with pressure that results in at least 5 ETTs in the 0-40 GPa pressure regime. The first two ETTs can be directly inferred from Figure~\ref{fig:bands} (and the right panels in Figure \ref{fig:fermi}) as arising from bands crossing the Fermi level along the symmetry directions. However the other three transitions cannot be seen in the projection of the BZ chosen in Figure 1. For example the third ETT is due to band crossing in the LM direction. Hence we used the FS plots to illustrate them in Figure~\ref{fig:fermi}. The changes in the Fermi surface occur due to the contributions from each of the 4 bands shown in Figure~\ref{fig:bands}. The pressures indicated are just above the transition points and the topological features involved in the transitions are indicated by the black arrows. The first transition occurring  around 2 GPa is the linking of the ``arms of the monster" as discussed by  Novikov {\it et al}~\cite{novikov_anisotropy_1999}. Here, bands 1 and 2 that  are degenerate along LHK, move off the FS at K and a part of the M$\Gamma$ branch  of band 2 crosses the FS. The second transition occurs around 7-8 GPa and involves the appearance of needle-like pieces originating  at the K point in the Brillouin zone (BZ) and stretched along the HK  branch of band 3, also discussed by Novikov {\it et al}~\cite{novikov_anisotropy_1999}. All further transitions mentioned in this work are entirely new and have not been predicted theoretically before. The third transition involves changes in topology of the FS from contributions from bands 3 and 4 along the LM branch in the BZ. This ETT occurs around 13 GPa.  In fact, band 4 completely moves above the Fermi surface and does not contribute to it further. The actual transition occurs along a non-standard line in the BZ. Hence, in order to help visualize the origin of this transition, we have provided  projections of the FS contribution from bands 3 and 4 on the $\left[220\right]$ plane of the BZ in the relevant pressure range in Figure~\ref{fig:fs_2d}. Band 3 has a butterfly-shaped contribution around the L point while band 4 has a diamond-shaped contribution around the same point. The central parts of both these features vanish between 13 and 14 GPa in the third transition. The next topological change occurs around 17-18 GPa  where the contributions around the LM branch from band 3 also vanish. In particular, it is the region corresponding to the ``wings of the butterfly" depicted in Figure~\ref{fig:fs_2d} that ceases to contribute to the FS. To further illustrate the onset of these two transitions we have also given band structure plots corresponding to both the third and fourth transitions along LM as well as line parallel to LM originating from a point slightly shifted from L (L$^+$) in the $\left[220\right]$ plane (see Figure~\ref{fig:ett34_1d}).  Along LM both bands 3 and 4 are roughly parabolic about L and move across the FS as the pressure increases beyond 13 GPa. However, along the shifted line  band 3 develops a negative curvature at L$^+$ and, hence, crosses the FS twice resulting in the wing feature mentioned above. As we move away from $\left[220\right]$ parallel to this line the wing feature persists for pressures beyond 13 GPa eventually vanishing around 17 GPa and resulting in the fourth transition.  A fifth transition occurs around 28 GPa along the HK branch of band 2 where the top and bottom parts of the FS contribution become detached from the center. In addition to these at least one other transition occurring beyond 40 GPa can be  anticipated from band 2 when the central part vanishes from the FS. However, we have not followed this transition in the present work.

\subsection{Elastic constants}

\begin{figure*}[t!]
\begin{center}
\includegraphics[width=7.0in,clip=]{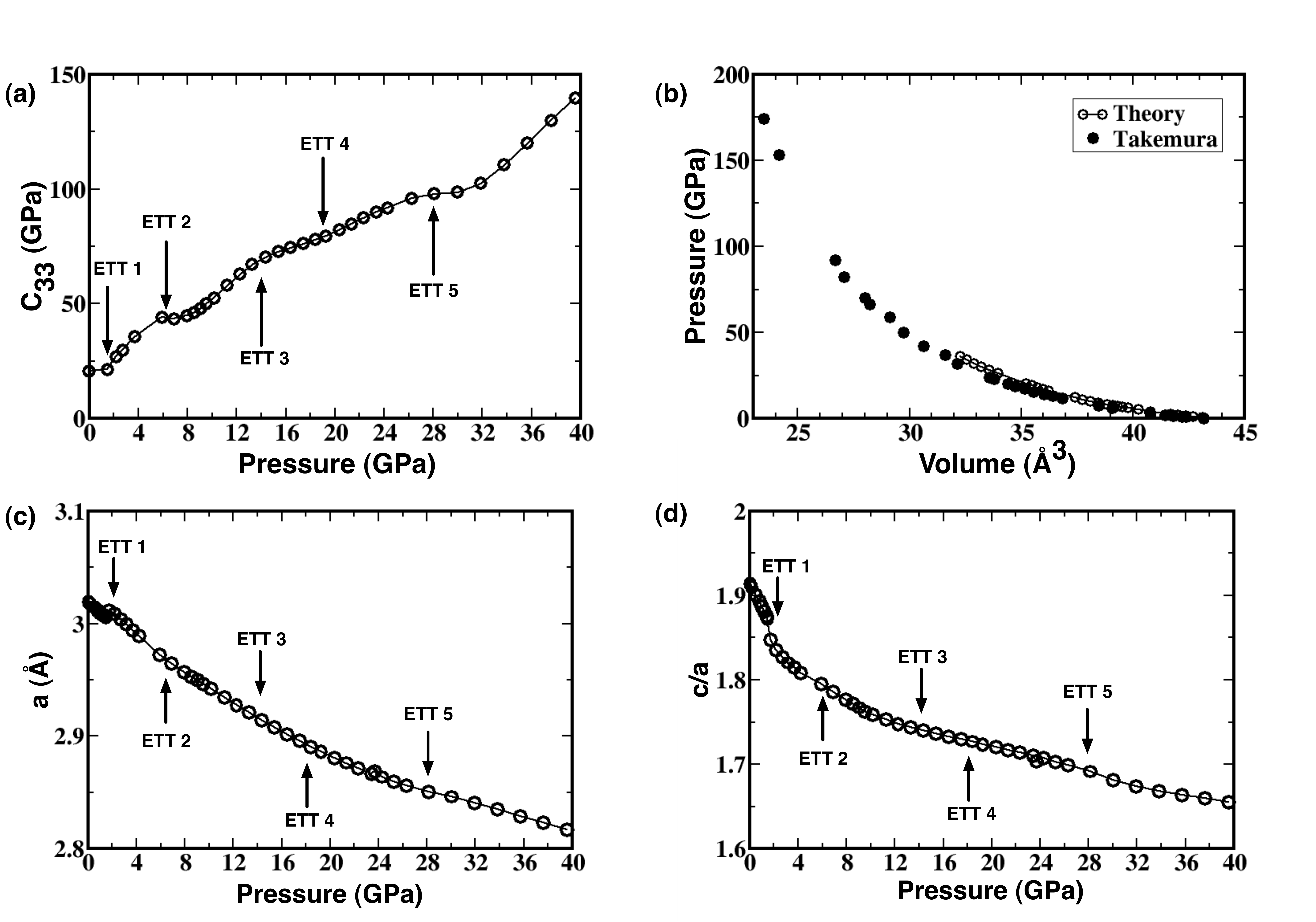}
\end{center}
\caption{(a) Computed dependence of the elastic constant $C_{33}$ on pressure indicating anomalies corresponding to the ETTs,  (b) calculated P-V isotherm at 300K compared with the experimental data of Ref. 6, (c) Variation of lattice parameter with pressure in hexagonal Cd, and (d) variation of the axial ratio $c/a$ with pressure in hexagonal Cd. The arrows indicate the location of the electronic topological transitions as predicted by the topological changes in the Fermi surface.
}
\label{fig:eos}
\end{figure*}

We also computed the elastic constant $C_{33}$ as an indicator of  the location of various ETTs.  The calculations of $C_{33}$  involved computing the 0~K free-energies $F(c,a) = E(c,a) + pV$ along the epitaxial Bain path~\cite{qiu_structural_2004} and obtaining the second derivative at the equilibrium structure at pressure {\it p} by a quadratic fit. Figure~\ref{fig:eos}a shows the dependence of $C_{33}$ on pressure thus obtained. It is clearly seen that anomalies in this dependence parallel the observation of ETTs in the Fermi surface plots in Figure~\ref{fig:fermi}. The location of the ETTs is perhaps not precise for one or both of two reasons: the Brillouin-zone sampling is insufficient or the variation of the elastic constants is effected by the rearrangements in the energy spectrum occurring around an ETT~\cite{sinko_effect_2005}.

\subsection{Comparison with experimental isotherm}

The difficulty in detecting ETTs under pressure motivated us to use first-principles theory to assess whether experimentally observed anomalies can  be correlated with electronic transitions. Hence the contact with the experimental data on Cd was made through total-energy calculations obtained as a function of volume to   obtain the 0 K pressure.  We added the zero-point vibrational and lattice-thermal contributions to the cold pressure at each volume for comparison with the measured 300~K isotherm for Cd by  Takemura~\cite{kenichi_structural_1997}, in accordance with the procedure illustrated in Ref. 28.

We obtained an equilibrium volume of 45.624~\AA$^3$, which is within 6\% of the experimental value of 43.159~\AA$^3$.  This is in accordance with the fact that GGA potentials typically yield high values for the equilibrium volume.  The value of the bulk modulus, its pressure derivative and the Debye temperature are 47.5 GPa, 5.3 and 373.3 K, respectively.  The P-V equation of state at 300~K is compared with the experimental data of Takemura in Figure~\ref{fig:eos}b. Good agreement is seen between the computed and experimental data although the theoretical curve  becomes slightly stiffer at high compressions which is consistent with the slightly higher value of the bulk modulus obtained in our calculations compared to the experimental value of 46 GPa. While we notice no anomaly in the 300 K P-V isotherm, the variation of lattice parameter $a$ and axial ratio $c/a$ as a function of pressure at 0 K do reveal discontinuities around 2, 8, 12, 18 and 28 GPa (see Figures~\ref{fig:eos}c and \ref{fig:eos}d).
The x-ray diffraction studies of Pratesi {\it et al}~\cite{pratesi_anomalies_2005} on Cd using a different pressure transmitting medium than Takemura for the trends of lattice parameters and the axial ratio c/a as a function of pressure reveal distinct anomalous slope change at V/V$_0$ =0.86 (P$\sim$8 GPa).

\section{Conclusions}

The results of band structure calculations and resulting Fermi surface plots reveal unambiguously the presence of   five ETTs upto 30~GPa with the possibility of one more transition beyond 40~GPa. These ETTs correlate with anomalies in the variation of   lattice parameter  $a$ and $c/a$ as well as the elastic constant $C_{33}$ at 0 K.  The reliability of these predictions is checked by comparing the theoretical and experimental equation of states  at room temperature. Our work indicates that the pressure variation of elastic constant shows strong anomalies near ETTs. The fact that  the 0 K structural parameters  $a$ and $c/a$ display  anomalies close to the pressures where ETTs occur suggests it is desirable to have their measurements at  low temperature in close intervals of pressure. However there have been difficulties in the past in achieving the required precision in such measurements. Similarly, there exist  difficulties in the measurement of elastic constants at high pressure in metallic  solids. It is possible that ETTs could be induced by a magnetic field but the field strength required in metallic solid  is beyond current technical possibilities~\cite{kozlova_magnetic-field-induced_2005, koudela_lifshitz_2006,blanter_theory_1994}. Thus low temperature, high pressure  resistivity and  thermo-electric power  measurements are the most  promising routes to observe ETTs~\cite{blanter_theory_1994}. We hope that our studies will provide sufficient impetus to various experimental efforts aimed at direct detection of ETTs.

\bibliography{references}
\bibliographystyle{apsrev4-1}

\end{document}